\documentclass[reprint,aps,amsmath,amssymb]{revtex4-1}

\usepackage{graphicx}
\usepackage{dcolumn}
\usepackage{bm}
\usepackage{hyperref}

\usepackage{amssymb,euscript,latexsym,amsmath}
\usepackage{graphicx}
\usepackage[dvips]{epsfig}
\usepackage{color}
\usepackage{epstopdf}
\usepackage{setspace}
\usepackage{subfigure}

\newcommand{\abs}[1]{\left| #1 \right|} 

\newcommand{\conj}[1]{\overline{#1}}
\newcommand{\real}[1]{\mathrm{Re} \left[ #1 \right]}

\newcommand{\eee}[1]{\mathrm{e}^{ #1 }}
\newcommand{\ii}{\mathrm{i}} 

\begin{document}

\title{Density shock waves in confined microswimmers} 

\author{Alan Cheng Hou Tsang and  Eva Kanso} \thanks{corresponding author: kanso@usc.edu}
 \affiliation{Aerospace and Mechanical Engineering, University of Southern California, Los Angeles, CA 90089}

\begin{abstract}
Motile and driven particles confined in microfluidic channels exhibit interesting emergent behavior from propagating density bands to density shock waves.  A deeper understanding of the physical mechanisms responsible for these emergent structures is relevant to a number of physical and biomedical applications. Here, we study the formation of density shock waves in the context of an idealized model of  microswimmers confined in a narrow channel and subject to a uniform external flow. Interestingly, these   density shock waves exhibit a transition from `subsonic' with compression at the back to `supersonic' with compression at the front of the population as the intensity of the external flow increases.
This behavior is the  result of a non-trivial interplay between hydrodynamic interactions and geometric confinement, and is confirmed by a novel quasilinear wave model that properly captures the dependence of the shock formation on the external flow. 
These findings can be used to guide the development of novel  mechanisms for controlling the emergent density distribution and average population speed, with potentially profound implications on various processes in industry and biotechnology such as the transport and sorting of cells in flow channels.

\begin{description}
\item[PACS numbers]
47.63.Gd,  87.18.Hf, 05.65.+b
\end{description}

\end{abstract}

\maketitle

\par\noindent
The transport of micro-particles in flow channels occurs in a variety of 
industrial and natural processes. Examples include 
filtration~\cite{hulin:1990},
 colloid deposition~\cite{deegan:nature1997a}, 
 droplet-based microfluidics~\cite{beatus:prl2009a}, blood cells in microflows \cite{popel:arfm2005a}, 
sperm cells in the fallopian tube~\cite{riffell:jeb2007a}, 
and pathogens in the urinary tract~\cite{mulvey:pnas2000a}. The ability to control the density distribution and group velocity
of micro-particles in flow would therefore have numerous applications in
physics and biology. 
This letter analyzes, via discrete particle simulations  and macroscopic models,  
the emergent patterns in populations of motile particles driven by an external flow in a rectangular micro-channel.

The dynamics of  single and many-swimmer systems are typically studied in unconfined settings~\cite{dunkel:prl2013a, saintillan:crp2013a, marchetti:rmp2013a}. The effects of confinement to narrow flow channels are less well understood; 
see, e.g.,~\cite{hill:prl2007a, berke:prl2008a, zottl:prl2012a,figuero:sm2015a,ezhilan:jfm2015a}
and references therein.
Recent experiments on driven droplets~\cite{beatus:np2006a, beatus:prl2009a, beatus:pr2012a, 
desreumaux:prl2013a} and self-propelled colloids~\cite{bricard:nature2013a, bricard:nc2015a}
in quasi two-dimensional (\textit{Hele-Shaw} type) channels show the emergence of 
traveling density waves, including density shocks at the wave front. 
Similar observations are reported in simulations of confined self-propelled
particles~\cite{lefauve:pre2014a}, albeit with density shocks forming at the back of the wave.
The primary mechanism responsible for the emergence of these density waves
is attributed to hydrodynamic interactions (HI) among the confined particles~\cite{bricard:nature2013a, lefauve:pre2014a}.
Confinement in \textit{Hele-Shaw} type geometries induces a distinct hydrodynamic signature; the particle far-field disturbance 
is that of a source dipole irrespective of the details of the particle transport 
mechanism, driven or self-propelled, pusher or puller~\cite{beatus:prl2009a, 
beatus:np2006a,beatus:pr2012a, desreumaux:epje2012a, brotto:prl2013a}. 
Additional confinement to a narrow channel does not alter the nature of hydrodynamic 
interactions but imposes impenetrability conditions on the lateral boundaries 
of the channel.  These effects are properly accounted for in the model used in~\cite{lefauve:pre2014a},
and, thus, do not explain the discrepancy  in the location of the shock 
formation between the simulations~\cite{lefauve:pre2014a} 
and experiments~\cite{beatus:pr2012a,bricard:nature2013a}.
Here, we show that  the background flow and geometric confinement coupled with the HI among the swimmers and their motility properties lead to a rich variety of emergent behaviors, consistent with experimental  evidence but unobserved in simulations before. Most notably, we show the emergence of density shocks that transition from  `subsonic' to `supersonic' as the intensity of the background flow increases. We conclude by demonstrating that this richness can be exploited to control the density distribution and collective speed of the microswimmers.

\begin{figure}[!t]
\centerline{\includegraphics[width=0.475\textwidth]{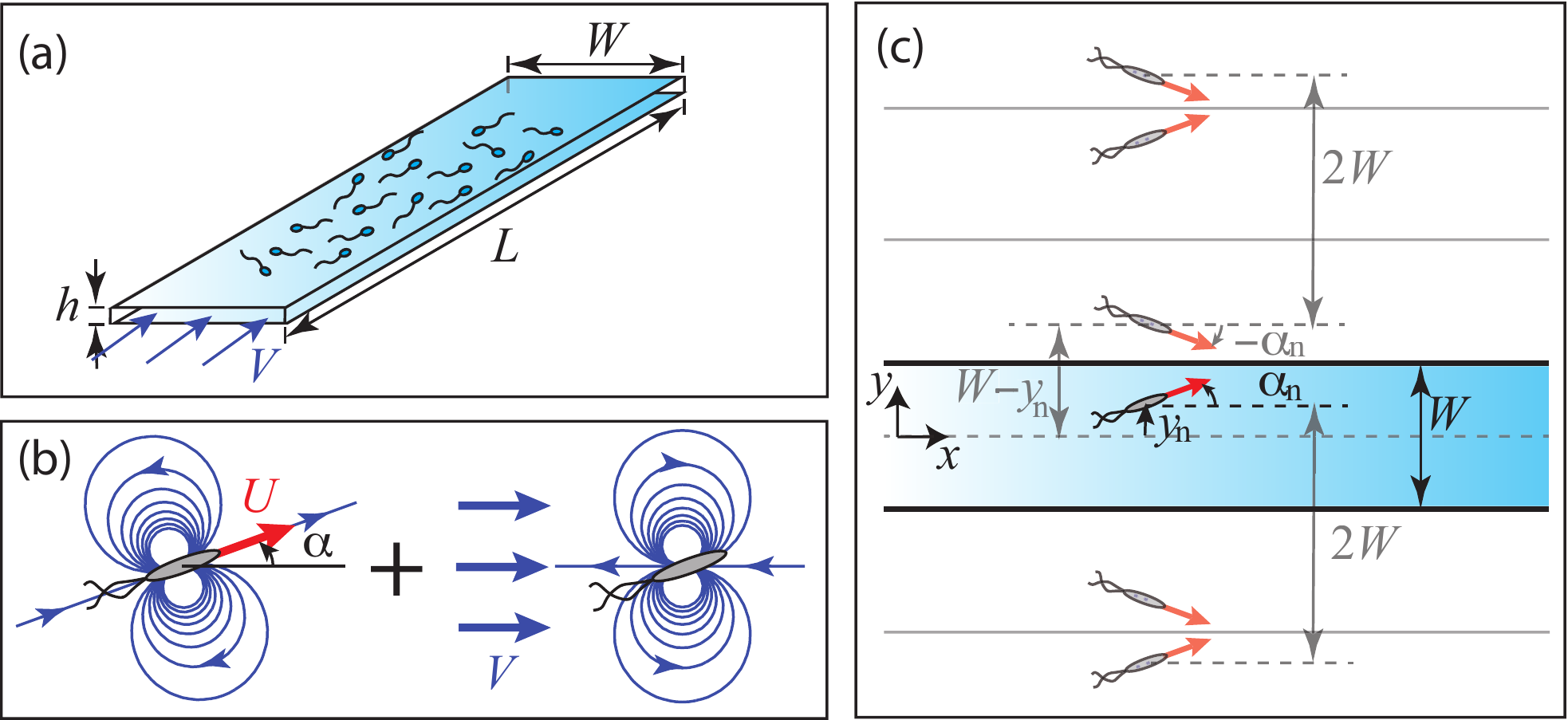}}
\caption[]{(a) Schematic of microswimmers in a Hele-Shaw rectangular channel. (b) 
Dipolar far-field flows induced by the self-propelled motion and the background flow. 
(c) Sidewall confinement is accounted for using an infinite image system. } 
	\label{fig:schematic}
\end{figure}

\begin{figure*}[!t]
\centerline{\includegraphics[width=1\textwidth]{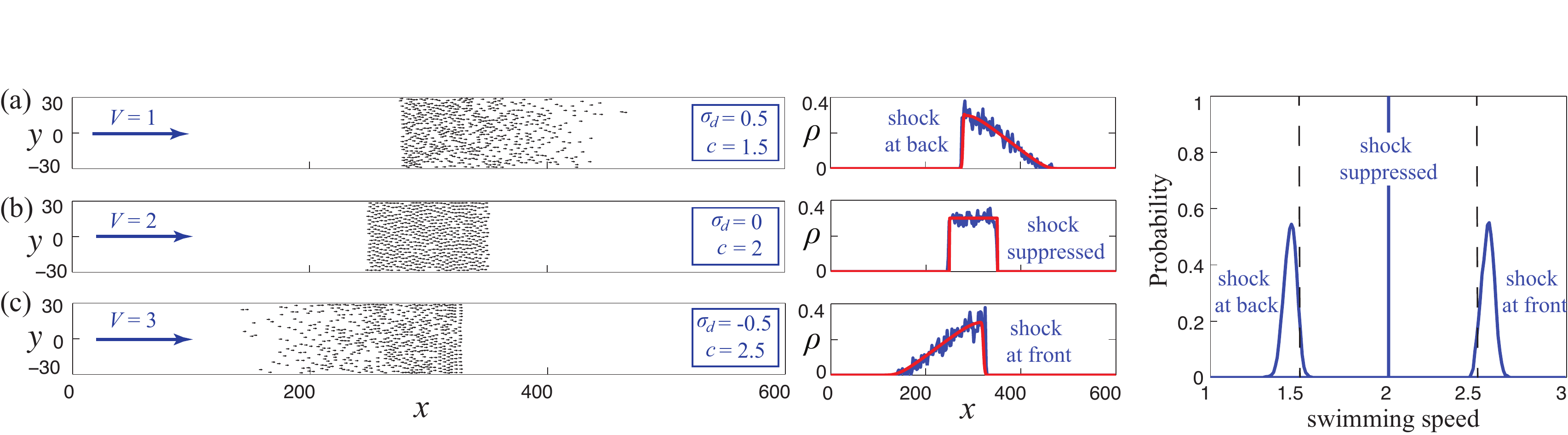}}
\caption[]{Density shock waves in confined microswimmers. Left: (a)  for $V < c$, compression shock wave forms at the back; (b)   for $V=c$, shock is suppressed; (c) for $V>c$, compression wave forms at the front.  The `speed of sound' of the medium ahead  is $c = U + \mu V$.  Middle: the density $\rho$ is the local area fraction computed as in~\cite{beatus:prl2009a}. Simulations results of the quasi 1d continuum model are superimposed in red. Right: the probability distribution of swimming speeds in the discrete simulations are shown in solid lines, and the speed of sound of the medium ahead is indicated in dashed lines.
Parameter values are: $W=60$, $\Phi_A=0.3$, $\nu=1$, snapshots taken at $t = 1000$.}
	\label{fig:shockwave}
\end{figure*}  

We briefly review the model of $N$ microswimmers in \textit{Hele-Shaw} confinement presented 
in~\cite{tsang:pre2014a,tsang:pre2015a} based on the work in~\cite{desreumaux:epje2012a,brotto:prl2013a} . 
Namely, one has
\begin{equation}
\begin{split}
\label{eq:eom}
  	\dot{\conj{z}}_n  = U \eee{-\ii \alpha_n}+ \mu \conj{w}(z_n)+ v_n, \quad
  	\dot{\alpha}_n  =  \real{\nu \conj{w} \ii  \eee{\ii \alpha_n}} 
\end{split}
\end{equation}
Here, $z_n$ denotes the swimmer's position in the complex plane $z = x+\ii y$ $(\ii = \sqrt{-1})$ and $\alpha_n$ its orientation, $n=1,\ldots, N$. The overline $\conj{(\cdot)}$ and the operator Re$[\cdot]$ denote the complex conjugate and the real part, respectively.   Each swimmer has a self-propelled speed $U$ and is advected by the local flow velocity $w(z)$. All variables are non-dimensional 
using the characteristic length and time scales dictated by the size and speed of the individual swimmer.  $\mu$ and $\nu$ are translational and rotational mobility coefficients whose values depend on the geometric properties of the swimmer and its flagellar activity: $\mu$ 
is in the range $0< \mu<1$ while $\nu$ is either positive or negative depending on the head-tail asymmetry~\cite{brotto:prl2013a} or flagellar activity~\cite{tsang:pre2014a}. 
We use a collision avoidance mechanism $v_n$ based on the repulsive part of the Leonard-Jones potential. These near-field interactions decay rapidly outside a small excluded area centered around $z_n$, thus ensuring that the order of the far-field HI is preserved.

To close the model in \eqref{eq:eom}, we need 
to evaluate the velocity field $w(z)$ created by $N$ confined swimmers in uniform background flow, 
(Fig.~\ref{fig:schematic}a). The self-propelled motion of a swimmer $z_j$ induces a dipolar velocity field $\conj{w}_s(z-z_j) = a^2U e^{i\alpha_j}/(z-z_j)^2$ which points in the swimming direction, whereas the background flow induces another dipolar field $\conj{w}_b(z-z_j) = -a^2(1-\mu)V/(z-z_j)^2$ which points in the  direction opposite to the background flow, irrespective to the orientation of the swimmer (see Fig.~\ref{fig:schematic}b and Ref.~\cite{desreumaux:epje2012a}).  Here, $a$ and $U$ are the effective radius and self-propelled speed of the swimmer, normalized to 1, and $V$ is the velocity of the background flow. This model is similar to the one used in~\cite{lefauve:pre2014a}. However, the fact that $w_b$  is proportional to $(1-\mu)$ was overlooked in~\cite{lefauve:pre2014a}, thus failing to correctly capture the dipolar field induced by the background flow. The velocity field $w(z)$ created by $N$ swimmers in uniform background flow can be written as a superposition of $w_s$ and $w_b$.

The impenetrability conditions at $z = \pm \ii W/2$ due to the sidewall confinement are accounted for using the method of images. Basically, to each swimmer at $z_i=x_i+ \ii y_i$ correspond two periodic lattices, of period $2\ii W$, of image dipoles located at $z_n=x_n+ \ii y_n$ and $z_n=x_n+\ii (W-y_n)$, respectively (Fig.~\ref{fig:schematic}c). To calculate the velocity field $\conj{w}$ due to $N$ swimmers and their image system, one has to evaluate infinite sums of terms that decay as $1/z^2$.   An approximate numerical solution is given in~\cite{lefauve:pre2014a}. Here, we derive an exact analytical solution in terms of hyperbolic functions, 
\begin{equation}
\label{eq:wallvelocity}
	\begin{split}
	\conj{w}(z_n) &= V + \Bigl(\frac{\pi a}{2W}\Bigr)^2 \Biggl\{\sum_{j \neq n}^N  \sigma_j \text{csch}^2 \Bigl[{\frac{\pi}{2W}(z_n-z_j)}\Bigr] \\[-2ex]
	& \hspace{0.5in}- \sum_{j=1}^N  \conj{\sigma}_j \text{sech}^2 \Bigl[{\frac{\pi}{2W}(z_n-\conj{z}_j)} \Bigr] \Biggr\}
	\end{split}
\end{equation}
where $\sigma_j= U \eee{\ii \alpha_j}- (1-\mu)V$ and $a^2\sigma_j$ indicates the strength and orientation of the dipole induced at $z_j$.
%
\begin{figure}[!t]
\centerline{\includegraphics[width=0.475\textwidth]{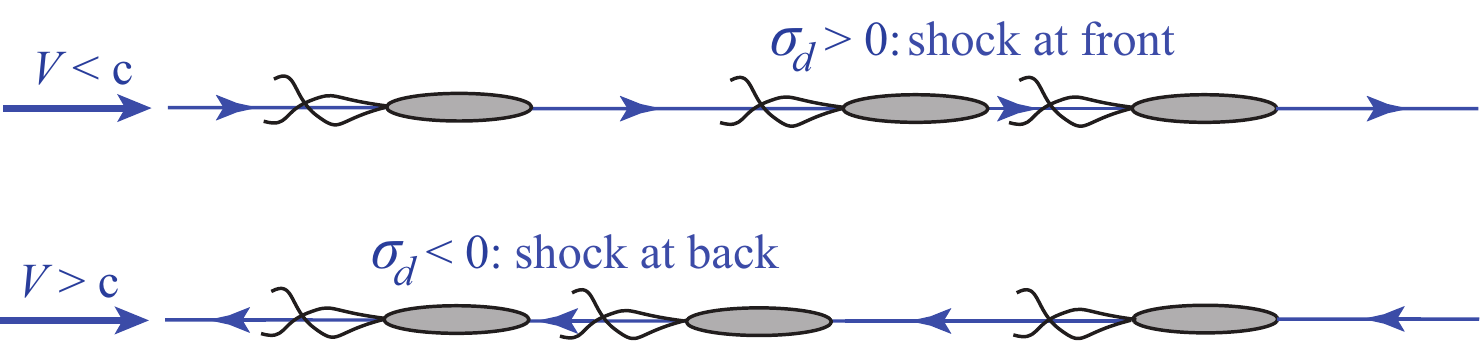}}
\caption[]{1d model does not capture correctly the shock wave formation. It results in compression shock at the front for $V<c$ and at the back for $V>c$, as opposed to the 2d system which exhibits compression at the back for $V<c$ and front for $V>c$ (Fig.~\ref{fig:shockwave}).}
	\label{fig:1Dsystem}
\end{figure}  
%

We examine the emergent behavior of populations of $N$ swimmers that initially fill the whole channel width and a segment $\ell$ of the channel length and are homogeneously placed and randomly orientated. Periodic boundary conditions are imposed at the two ends of the channel. We  set $\mu=0.5$. 
When the swimmers are subject to a sufficiently strong background flow, they quickly reorient and align in either the same (for $\nu>0$) or opposite (for $\nu<0$) direction  to the background flow. Basically, the background flow suppresses the orientation dynamics of the swimmers as noted in~\cite{lefauve:pre2014a}. All $\sigma_j$ reach, in finite time, a developed value $\sigma_d=pU-(1-\mu)V$, where $p=\textrm{sgn}(\nu)$ and sgn is the signup function.  It turns out that $\sigma_d$, which we refer to as the effective polarization parameter, is a key parameter for understanding the emergent structure of the swimmers.

The swimmers self-organize into a shock wave structure where, for $\nu>0$, the location of the compression wave depends on the strength of the background flow $V$, which in turn dictates the `speed of sound' of the medium given by $c= U + \mu V$ (see Supp. Mat. and Supp. Movie 1).
For a weak background flow $V<c$, the effective polarization $\sigma_d$ of the swimmers  is dominated by their self-propelled motion so that $\sigma_d>0$. The emergent structure is characterized by a compression wave at the back and a rarefaction or expansion wave at the front (Fig.~\ref{fig:shockwave}a). The average population speed is smaller than $c$, thus this shock wave can be termed as `subsonic'.  If we tune the background flow properly such that the polarizations induced by the self-propelled motion and background flow cancel out such that $\sigma_d=0$, the shock wave is suppressed and the swimmers evolve as a  uniform cluster traveling at the speed $V$ (Fig.~\ref{fig:shockwave}b). When the background flow is strong $V>c$, it dominates the swimmers' effective polarization $\sigma_d<0$; the density shock forms at the front rather than at the back (Fig.~\ref{fig:shockwave}c), and the average population speed is greater than $c$, thus the term `supersonic.' Note that for  $\nu<0$, the swimmers align in the opposite direction to the background flow. The developed polarization parameter $\sigma_d$ is always negative, and thus the swimmers develop a shock wave structure similar to Fig.~\ref{fig:shockwave}c that could travels upstream or downstream depending on the value of $V$. 

The density shock formed in Fig.~\ref{fig:shockwave}c is reminiscent to the experimental results of~\cite{beatus:prl2009a} for driven droplets in a Hele-Shaw channel and of the propagating density bands shown in~\cite[supp. videos]{bricard:nature2013a} for self-propelled colloids. The propagating density fronts in the latter system and their similarity to our results are surprising; however,
the agreement with~\cite{beatus:prl2009a} is consistent with the intuition that active swimmers  behave like passive particles when subject to strong background flow.


We now  consider the simple picture of three swimmers  perfectly aligned parallel to the $x$-axis. 
One can easily see that for $V<c$  and $\sigma_d>0$,  HI  push the swimmers forward such that the middle swimmer is always faster leading to compression at the front, whereas for $V>c$ and $\sigma_d<0$, HI hinder the swimming motion and the middle swimmer is always slower leading to compression at the back (Fig.~\ref{fig:1Dsystem}). This argument can be generalized to $N$ swimmers. Thus, in 1d systems, the density shock forms at the front of the swimmers when the background flow is weak and  at the back when the background flow is strong.
 This prediction,  illustrated in Fig.~\ref{fig:1Dsystem}, agrees with the density shocks seen in the 1d experiments of  driven microfluidic droplets~\cite{champagne:sm2011a}, and 
is exactly opposite to the 2d experiments~\cite{beatus:prl2009a} and simulations presented in Fig.~\ref{fig:shockwave}. The contradiction reveals the importance of 2d HI and the complexity of the mechanism responsible for the formation of density shock waves in populations of microswimmers in confined 2d channels.

Density shocks emerge as a result of the interplay between 2d hydrodynamics, sidewall confinement, and the initial distribution of swimmers into a segment $\ell$ of the full channel. For $W =\infty$ and swimmers homogeneously filling the whole space, no shock develops due to the radial symmetry of the dipolar field at each swimmer. When sidewall confinement is introduced, it serves as a mechanism to break the radial symmetry of the dipolar field and creates a non-zero average hydrodynamic interaction acting on the swimmers. The initial uniform distribution of swimmers into a segment $\ell$ of the full channel length creates a spatial gradient in the HI in the $x$-direction, which contributes to triggering the instability of the initially uniform swimmer distribution.

\begin{figure}[!t]
\centerline{\includegraphics[width=0.5\textwidth]{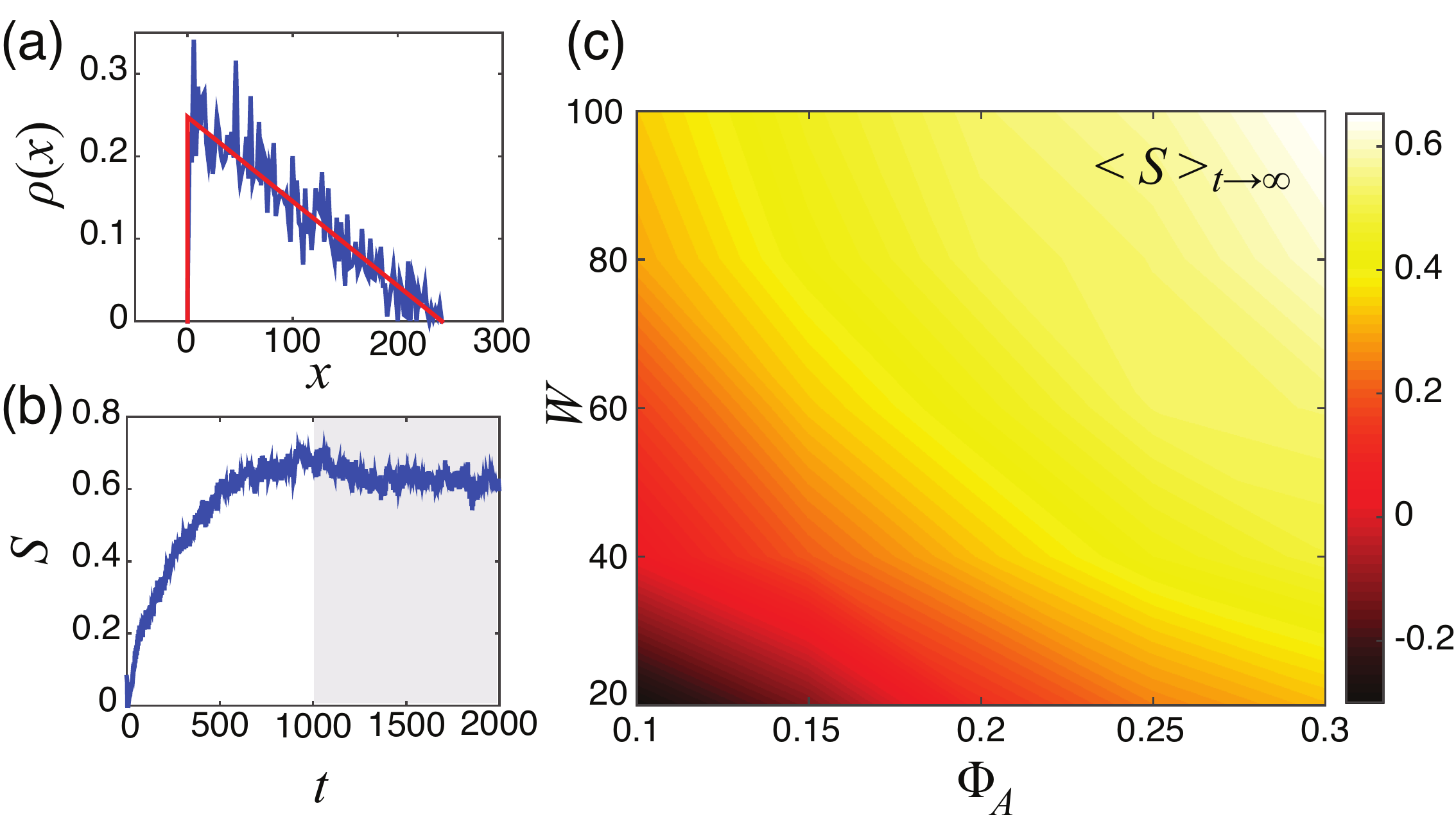}}
\caption[]{(a) Shock order parameter is defined by comparing the actual density profile to  a triangular density profile, shown here at $t=2000$ for $W=80$ and $\Phi_A=0.3$. (b) Evolution of the shock order parameter for the case in $(a)$.  (c) Mean value of the shock order parameter as a function of channel width $W$ and swimmers' area fraction $\Phi_A$.}	\label{fig:shockorder}
\end{figure}

We numerically investigate the effect of 2d interactions on the shock wave formation by fixing $\nu=1$ and $V=1$ (i.e., $\sigma_d=0.5$) and varying the channel width $W$ and the initial local area fraction
$\Phi_A= \pi a^2N/(W \ell)$.
To obtain the desired value of $\Phi_A$, we set $\ell=100$ and vary $N$. For each set of parameters, we run multiple trials corresponding to different sets of initial conditions taken from a uniform probability distribution function (Monte-Carlo type simulations). We capture the emergence of the density shock wave using the shock order parameter
\begin{equation}
\label{eq:shockparameter}
	S(t)=1-2 \sum_x\abs{\rho(x,t)-\rho_{\rm fit}(x,t)}/ \sum_x \rho_{\rm fit}(x,t)
\end{equation}
where $\rho(x)$ is the local density of the swimmers and $\rho_{\rm fit}(x)$ is a triangular fit of the density profile $\rho(x)$ (Fig.~\ref{fig:shockorder}a). $S = 1$ corresponds to a perfectly triangular shock and $S = 0$ to a rectangular homogenous profile. An example of the evolution of $S$ is depicted in Fig.~\ref{fig:shockorder}b. The dependence of the mean value $\langle S\rangle_{t\rightarrow \infty}$ averaged over all simulations on the parameter space ($W,\Phi_A$) is depicted in Fig.~\ref{fig:shockorder}c. It shows a continuous increase in  $\langle S\rangle_{t\rightarrow \infty}$ as $W$ and $\Phi_A$ increase which then levels off for large $W$ and $\Phi_A$. This plot provides two important pieces of information. First, it serves as a strong supporting evidence that 2d HI is important for the shock wave formation -- for a given $\Phi_A$,  an ordered shock is only formed when $W$ exceeds a critical value. Second, it shows that the density shock formation is robust over a large plateau of $W$ and $\Phi_A$.

To further elucidate the effect of the complex interplay between the 2d HI, sidewall confinement, and background flow on the density shock formation,  we develop a reduced continuum model that properly accounts for all these elements. 
Our model is distinct from the traffic flow model used in~\cite{lefauve:pre2014a} and Burgers' equation used in~\cite{beatus:prl2009a}. 
These equations oversimplify the effect of HI. They are only correct to leading order in the weak confinement limit with $a \ll W$ and are not capable of explaining the transition of shock formation from subsonic to supersonic as the background flow velocity $V$ increases. 
Here, we assume  the swimmers' density is homogenous in the $y$-direction and approximate the swimmers' number density as a 1d function $\rho_c(x,t)$. The continuum number density $\rho_c$ is related to $\rho$ via $\rho_c=\rho/(\pi a^2)$. Following a standard homogenization procedure, we develop a 1d continuum model (recall that $p = \textrm{sgn}(\nu)$)
\begin{equation}
\label{eq:continuum}
	\frac{\partial \rho_c}{\partial t} +  \frac{\partial}{\partial x} \Bigl[ \left(U p+\mu V+ \mu u\right) \rho_c \Bigr]=0.
\end{equation}
The 2d HI are captured by the term $u(x,t)$
\begin{equation}
\label{eq:integral}
	u(x,t)= \frac{1}{\!W\!-\!2a\!}\text{Re}\left[ \int \!\!  \!\!  \int  \!\!  \!\!  \int \conj{w}_c(z, z^{\prime})\rho_c(x^{\prime},t) dydy^{\prime}dx^{\prime} \right]
\end{equation}
where  $x^\prime$ is integrated from $0$ to $L$ and $y$, $y^\prime$ from $-W/2+a$ to $W/2-a$, while the  dipolar flow field $\conj{w}_c$  is given by
\begin{equation*}
\label{eq:continuumvel}
\begin{split}
	\conj{w}_c  =  \sigma_{d}\Bigl(\frac{\pi a}{2W}\Bigr)^2 \Bigl\{ \text{csch}^2 \Bigl[{\frac{\pi}{2W} (z-z^{\prime})} & \Bigr] -  \\[-1ex]
	& \text{sech}^2 \Bigl[{\frac{\pi}{2W} (z - \conj{z}^{\prime})} \Bigr]\Bigr\}
	\end{split}
\end{equation*}
We account for the steric repulsion between the swimmers via the excluded area $\abs{x-x^\prime}^2+\abs{y-y^\prime}^2<4a^2$, inside which the dipolar contribution is removed. The effect of the excluded area is also reflected in the upper and lower limits of the $y$-integral in \eqref{eq:integral}, indicating that no swimmers can penetrate the sidewall boundaries.
We solve the continuum equation \eqref{eq:continuum} numerically 
with periodic boundary conditions. 
The numerical solutions, superimposed in red in {Figure~\ref{fig:shockwave}, show excellent agreement between the discrete particle simulations and the continuum model, including successfully capturing the transition in the location of the shock formation as  the background flow increases and the time scale of wave dispersion (see Supp. Movie 1).

This letter provides the first systematic study of the emergence of compression shock waves in populations of mircoswimmers driven in narrow  flow channels. Our results are consistent with experimental observations on driven droplets and self-propelled colloids,~\cite{beatus:np2006a,beatus:prl2009a,bricard:nature2013a, bricard:nc2015a}, but go beyond these observations to report an unobserved transition from  `subsonic' to `supersonic' compression shocks as the intensity of the background flow increases. The physical mechanisms underlying this transition are elucidated via discrete simulations and a continuum model that properly captures the effects of HI and geometric confinement. A rigorous analysis of the onset and propagation of instability in the continuum model is deferred to future work.

This physical understanding of the emergent behavior and its dependence on the swimmers' motility properties and intensity of background flow can be exploited to devise novel mechanisms for controlling populations of microswimmers in externally-driven flow channels. Two species of microswimmers of different motility properties subject to the same background flow $V$ get segregated in space in finite time (see Supp. Movie 2). This is particularly useful for biomedical applications such as sorting of cells in flow channels.  Another compelling scenario
is to control the throughput, i.e., the speed of a cell population through a flow channel, by careful manipulation of the background flow $V$.  One could also control the profile of the population density distribution. For example, a Gaussian density profile can be obtained and maintained from an initially uniform distribution by properly tailoring a time-dependent background flow $V$ (see  Supp. Movie 3). These results, albeit in the context of a simplified model, have profound implications on developing a physics-based, quantitative framework for the design and control of particle-laden flows in microfluidic channels.

\bibliography{reference}

\end{document}